\documentclass[sigplan,screen]{acmart}
\AtBeginDocument{%
  }

\renewcommand\footnotetextcopyrightpermission[1]{}
\setcopyright{none}

\usepackage{amsmath}

\usepackage{booktabs}
\usepackage{makecell}
\usepackage{xcolor}

\definecolor{goodgreen}{RGB}{0,120,70}
\newcommand{\good}[1]{\textcolor{goodgreen}{#1}}
\usepackage{tabularray}
\UseTblrLibrary{booktabs}
\usepackage{booktabs}
\usepackage{caption}
\usepackage[table]{xcolor}
\usepackage{algorithm}
\usepackage{algpseudocode}
\usepackage{makecell}
\usepackage{multirow}
\usepackage{tabularray}
\UseTblrLibrary{booktabs}
\usepackage[table]{xcolor}
\definecolor{mainblue}{RGB}{47,111,187}
\definecolor{mainpink}{RGB}{217,95,141}
\definecolor{softblue}{RGB}{238,244,255}

\begin{document}

\title{StateFlow: Sequence Pipeline Parallelism for Long-Context Modeling with Linear Recurrence}

% Author information for the arXiv preprint.
\author{Wenxuan Zhao}
\email{zhaowenxuan866@gmail.com}
\affiliation{%
  \institution{Tsinghua University}
  \city{Beijing}
  \country{China}}

\author{Yingfa Chen}
\email{chenyingfa1999@gmail.com}
\affiliation{%
  \institution{Tsinghua University}
  \city{Beijing}
  \country{China}}

\author{Xu Han}
\email{han-xu@mail.tsinghua.edu.cn}
\affiliation{%
  \institution{Tsinghua University}
  \city{Beijing}
  \country{China}}

\author{Wenjing Han}
\email{wenjinghan@tsinghua.edu.cn}
\affiliation{%
  \institution{Tsinghua University}
  \city{Beijing}
  \country{China}}

\author{Tianbo Huang}
\email{mlbthtb@126.com}
\affiliation{%
  \institution{ByteDance Inc.}
  \city{Beijing}
  \country{China}}

\author{Zhiyu Li}
\email{lizhiyu.0@bytedance.com}
\affiliation{%
  \institution{ByteDance Inc.}
  \city{Beijing}
  \country{China}}

\author{Ao Sun}
\email{why1.2seed@gmail.com}
\affiliation{%
  \institution{Beijing University of Posts and Telecommunications}
  \city{Beijing}
  \country{China}}

\author{Jingheng Xu}
\email{xujh@himis-sz.cn}
\affiliation{%
  \institution{Hetao Institute of Mathematics and Interdisciplinary Sciences}
  \city{Shenzhen}
  \country{China}}

\author{Lin Gan}
\email{lingan@tsinghua.edu.cn}
\affiliation{%
  \institution{Tsinghua University}
  \city{Beijing}
  \country{China}}

\author{Guangwen Yang}
\email{ygw@tsinghua.edu.cn}
\affiliation{%
  \institution{Tsinghua University}
  \city{Beijing}
  \country{China}}

\renewcommand{\shortauthors}{Zhao et al.}

\begin{abstract}

Long-context training is increasingly important for large language models, and linear attention and state space models have become popular for improving long-context efficiency. However, efficiently parallelizing long-sequence training for recurrent and hybrid models remains challenging.

We present StateFlow, a sequence pipeline parallelism system for models with linear recurrence. StateFlow partitions each sequence into chunks and schedules their execution while propagating boundary states and gradients across chunks, thereby reducing activation lifetimes and improving training throughput. StateFlow further uses profile-guided nonuniform chunking to balance recurrence and softmax attention computation in hybrid models, and overlaps state transitions that expose limited parallelism with surrounding computation.
Applying StateFlow to models with up to 32B parameters and 256K context length, we achieve up to \(2.22\times\) throughput improvements and \(2.45\times\) memory reduction compared to conventional pipeline parallelism, enabling otherwise infeasible configurations.

\end{abstract}




\maketitle

\section{Introduction}

Long-context language models are important for applications such as long-document understanding, retrieval-augmented generation, and agentic workloads~\cite{longcontext_survey,rag,agent}. However, the quadratic computational cost of softmax attention and the linear growth of activation memory with the context length make long-context training slow and expensive.

% However, increasing context length significantly amplifies the memory footprint and execution latency of training.
An alternative to softmax attention is linear recurrence, including linear attention~\citep{gla} and state space models~(SSMs)~\citep{mamba}. Unlike softmax attention, linear recurrence compresses contextual information into a fixed-size state, resulting in linear time complexity and constant space complexity. Thus, linear recurrence is more efficient for long-context processing and has been incorporated into recent open-weight models such as Kimi K3~\citep{kimi_k3}, Nemotron 3 Ultra~\citep{nemotron3_ultra}, and Qwen3.6~\citep{qwen3_6}.

Despite these computational benefits, activation memory remains a primary
bottleneck in large-scale long-context training. Existing parallelism strategies only partially address it.
Data parallelism~(DP) is constrained by global batch size and does not reduce the per-device model weight storage, while tensor parallelism~(TP) incurs collective communication and depends on interconnection bandwidth~\cite{megatron_lm}. Context parallelism~(CP) reduces activation memory through sequence partitioning but incurs substantial intra-layer communication for cross-partition attention~\cite{ringattention,ulysses,sequence_parallelism}.
Pipeline parallelism~(PP) reduces per-device model memory, yet each pipeline task spans the full sequence, and schedules retain multiple such microbatches in flight, causing high peak activation memory~\cite{gpipe,dapple,megatron_lm}.
% These approaches therefore do not fully address the activation memory bottleneck of long-context linear attention training.

More recently, \textit{sequence pipeline parallelism}~(SPP) addresses this memory overhead by splitting the sequence into chunks and scheduling them as pipeline units~\cite{terapipe,seq1f1b}. Because this approach schedules at a smaller granularity, it reduces memory footprint compared with conventional PP and improves throughput by reducing bubbles.
However, it is still unclear how to apply SPP to linear recurrence models efficiently.

Softmax attention and linear recurrence have different cross-chunk dependencies.
% However, softmax attention and linear recurrence have different cross-chunk dependencies. 
In softmax attention, chunks need access to all previous key/value tensors, whereas recurrent chunks depend only on fixed-size boundary states, allowing more efficient computation. Yet these states impose strict dependencies that existing SPP systems do not handle efficiently. Therefore, we present \textbf{StateFlow}, a low-latency, state-aware SPP system for recurrent and hybrid models.

While linear recurrence allows each chunk to be computed more efficiently than softmax attention, applying SPP to linear recurrence introduces new challenges.
% Using chunks as scheduling units introduces new challenges. 
First, the chunks are not independent. Correct execution must preserve recurrent states across chunk boundaries, maintain boundary states from local convolution modules, and propagate state gradients during backward computation.
% Second, the recurrent state update forms a strict critical path across outer chunks. Although this state passing computation carries only a compact state, its GPU kernel often exposes limited parallelism compared with the dense tensor computations used to prepare chunk-level update matrices and produce chunk outputs. A dependency-preserving chunk schedule can still leave much of the state passing latency exposed.
Second, linear recurrence depends on a \textit{state transition} that exposes limited parallelism compared with dense tensor computations that prepare chunk-level update matrices and outputs. This causes low GPU utilization during state transition.
Third, many models incorporate both linear recurrence and softmax attention layers~\cite{kimi_linear,hybrid_linear_attention_done_right}, complicating sequence partitioning. Linear recurrence favors uniform chunks with regular state transition workloads, whereas softmax attention may benefit from nonuniform chunks because later chunks attend to longer prefixes and therefore require more FLOPs~\cite{terapipe,mepipe}.

StateFlow addresses these challenges with three mechanisms. First, a state-aware chunk schedule preserves all cross-chunk dependencies and releases each chunk's activations after its local backward computation. Second, StateFlow overlaps state transitions with independent dense computation in both forward and backward passes. Third, it searches over hybrid-aware partitioning candidates to balance recurrent and softmax attention workloads.

% Finally, for hybrid models, StateFlow optimizes the sequence partitioning scheme to account for the fact that attention computation of each chunk scales linearly with its position in the sequence.

We conduct experiments on Gated DeltaNet and Mamba-3
models~\citep{gated_delta_networks,mamba3} with up to 32B parameters. With activation
recomputation, StateFlow achieves up to \(2.2\times\) end-to-end speedup and
\(2.45\times\) peak memory reduction on recurrent models, and up to
\(2.22\times\) speedup and \(2.44\times\) peak memory reduction on hybrid
models. Without recomputation, it achieves up to \(1.79\times\) speedup and
\(2.54\times\) peak memory reduction. We will release the source code for reproducibility.

% This paper presents \textbf{StateFlow}, a sequence-level pipeline execution system for long context training of linear attention and hybrid attention models. The key observation is that long context linear attention needs sequence partitioning to reduce activation memory, and its computation structure makes such partitioning suitable for sequence level pipeline scheduling. Given the required boundary state from the preceding chunk, the computation of a linear attention chunk is determined by its local length rather than its position in the sequence. Equal length chunks therefore perform the same local computation and form regular pipeline units, unlike softmax attention where later chunks may involve more prefix dependent work. StateFlow schedules these outer chunks as pipeline units, keeps activations only for in flight chunks, and releases them after local backward computation, while preserving the recurrent boundary states, auxiliary boundary states, and backward state gradients required for correct training. As a result, StateFlow reduces peak activation memory without changing model semantics.

In brief, this paper makes the following contributions:

\begin{itemize}
    \item We design \textbf{StateFlow}, a dependency-preserving SPP approach that reduces linear recurrence activation memory and pipeline bubbles by pipelining sequence chunks with compact state transition.
    \item We extend StateFlow to hybrid models with an offline partition search that balances the workload of linear recurrence and softmax attention.
    \item We reduce the state transition utilization bottleneck with optimized kernels and overlapping them with neighboring computation in both forward and backward passes.
    % \item We implement StateFlow on A100 GPUs and evaluate it on linear attention and hybrid attention workloads, showing ...
\end{itemize}

\section{Background}
\label{sec:background}

\subsection{Modern Linear Recurrence}

Currently, most linear recurrence falls into two categories: linear attention and state space models~(SSM).
While these two kinds of architectures were developed based on different theoretical motivations, they can both be viewed as a kind of recurrent neural network~(RNN) with matrix-valued recurrent state and a linear \textit{update rule}. Below, we provide a formulation that encompasses both linear attention and SSM, while keeping the details relevant to this paper.
% Then, we describe their implementations that enable parallel training. Finally,  preliminaries for hybrid architectures and parallelization techniques for long-context training.

% \subsection{Linear and Hybrid Attention Execution}
% \label{sec:linear-hybrid-execution}

% Linear attention replaces the explicit prefix key/value cache of softmax
% attention with a compact recurrent state~\cite{efficient_attention_survey,gla,
% gated_delta_networks,kimi_linear}.

Let \(x_t \in \mathbb{R}^{d}\) and \(y_t \in \mathbb{R}^{d_v}\) denote the input and output representations at time step \(t\). The core computation of a linear recurrence layer can be written as
\begin{equation}
    S_t = S_{t-1}F_t + v_tk_t^\top,
    \qquad
    y_t = S_tq_t .
\end{equation}
Here, \(F_t \in \mathbb{R}^{d_k \times d_k}\), \(q_t, k_t \in \mathbb{R}^{d_k}\), and \(v_t \in \mathbb{R}^{d_v}\) are functions of \(x_t\). The two equations are known as the \textit{update rule} and the \textit{query rule}, respectively, and \(S_t\) is called the \textit{recurrent state}\footnote{Also called the \textit{hidden state} in some papers.}. For all models studied, \(S_0=0\).

To enable parallel training and reduce parameter count, \(F_t\) is usually structured. For Mamba-2~\citep{mamba2} and Mamba-3~\citep{mamba3}, \(F_t\) is a scalar-identity matrix; for Gated Linear Attention, it is a rank-1 matrix~\citep{gla}; for Gated DeltaNet~(GDN), it is a scalar-plus-low-rank matrix~\citep{gated_delta_networks}; and for Kimi Delta Attention~(KDA), it is a diagonal-plus-low-rank matrix~\citep{kimi_linear}.

\subsection{Chunkwise Parallelism for Linear Recurrence}
\label{sec:chunk-wise-parallelism}

% Directly using the recurrent formula to process the input tokens sequentially is prohibitively slow on modern hardware, which is optimized for parallel computation. In practice, for training and prefilling, linear recurrence implementations utilize a chunk-wise parallel formulation to enable parallel processing of input tokens while keeping the time complexity linear~(as opposed quadratic for softmax attention).
Efficient linear recurrence implementations process tokens in fixed kernel-level chunks rather than applying recurrence token by token~\citep{gla}. For a chunk of input representations \(X_i=[x_{iT_C}, \cdots, x_{(i+1)T_C -1}]\), where $T_C$ is the chunk size, we abstract its execution as three functions, ${Pre}, {ST}, {Out}$:
% To abstract away unnecessary details, chunk-wise parallel implementations partition a sequence of embeddings $X\in\mathbb R^{T\times d}$ into chunks of size $T_C$. Chunkwise implementations expose parallel computation across sequence chunks, while the recurrent state transition remains sequential along the chunk dimension. Let $X_{iT_C:(i+1)T_C},Y_{iT_C:(i+1)T_C}$ denote the input and output representations of the $i$-th chunk, the implementation can be formalized as:
\begin{equation}
\begin{aligned}
    \Omega_i
    &=
    Pre\!\left( X_i \right), \\
    \left(S_{(i+1)T_C}, V_i^{\mathrm{new}}\right)
    &=
    {ST}\!\left(S_{iT_C}, \Omega_i\right), \\
    Y_{i}
    &=
    {Out}\!\left(V_i^{\mathrm{new}}, S_{iT_C}, \Omega_i\right).
\end{aligned}
\end{equation}
where $\Omega_i$ is the set of data-dependent parameters used by state transition and output computation, while $V^\text{new}_i$ is auxiliary representation exposed by the recurrent operator for the delta rule (for SSMs, we have $V^\text{new}_i = \emptyset$), $Y_i$ is the output representation of this chunk.

\subsection{Hybrid Architectures}
\label{sec:hybrid-models}

Linear recurrence might underperform softmax attention on recall-intensive tasks~\citep{statex}. Thus, many long-context models use hybrid stacks with one softmax attention layer per \(3\)--\(7\) linear recurrence layers, which can achieve Transformer-level recall while retaining much of the efficiency benefit of linear recurrence~\cite{hybrid_attention,kimi_linear, hybrid_linear_attention_done_right}. When training such models with PP, the training system needs to account for the different complexities of the two kinds of layers to balance the workload of each device, in order to minimize bubbles and maximize hardware utilization.

\subsection{Parallel Training for Long Sequences}
\label{sec:parallel-training}

\begin{table}[!tbp]
    \centering
    \caption{
    Qualitative comparison of sequence pipeline parallelism methods.
    }
    \label{tab:related_parallel}

    \fontsize{8pt}{9pt}\selectfont
    \setlength{\tabcolsep}{1.8pt}
    \renewcommand{\arraystretch}{1.05}

    \begin{tabular*}{\columnwidth}{
        @{\extracolsep{\fill}}lcccc@{}
    }
        \toprule
        Method
        & \makecell{Communication\\volume}
        & \makecell{Activation\\memory}
        & \makecell{Balanced\\runtime}
        & \makecell{Linear\\recurrence} \\
        \midrule

        TeraPipe~\cite{terapipe}
        & \good{Low}
        & High
        & \good{Yes}
        & No \\

        Seq1F1B~\cite{seq1f1b}
        & \good{Low}
        & \good{Low}
        & \good{Yes}
        & No \\

        MEPipe~\cite{mepipe}
        & \good{Low}
        & \good{Low}
        & No
        & No \\

        SlimPipe~\cite{slimpipe}
        & High
        & \good{Low}
        & No
        & No \\

        \midrule

        \textbf{StateFlow (ours)}
        & \good{\textbf{Low}}
        & \good{\textbf{Low}}
        & \good{\textbf{Yes}}
        & \good{\textbf{Yes}} \\

        \bottomrule
    \end{tabular*}
\end{table}
Many large-scale training systems adopt sophisticated parallelism strategies to reduce per-device costs by partitioning computation and model states. Megatron-LM combines TP and PP, while ZeRO shards model states across data-parallel workers~\cite{megatron_lm,megatron_lm_2021,zero}. Pipeline schedules such as PipeDream, interleaved 1F1B, and Zero Bubble improve utilization by reordering microbatch execution~\cite{pipedream,megatron_lm,zerobubble_pipeline}. However, they still schedule full-sequence microbatches, whose activation lifetime and latency grow with the context length.

Long-sequence training often relies on sequence parallelism~(SP), which partitions activations and computation along the sequence dimension. DeepSpeed-Ulysses redistributes sequence and attention-head partitions through all-to-all communication, while Ring Attention circulates key/value blocks and overlaps communication with blockwise attention~\cite{ulysses,ringattention}. USP and LoongTrain combine multiple partitioning and communication schemes~\cite{usp,loongtrain}. For linear attention, LASP-2 and ZeCO distribute sequence computation through compact state communication~\cite{lasp2,zeco}. These methods provide intra-layer SP.

A complementary direction is sequence pipeline parallelism~(SPP). TeraPipe, Seq1F1B, SlimPipe, and MEPipe expose sequence segments as pipeline units to balance causal attention, reduce pipeline bubbles, or release activations earlier~\cite{terapipe,seq1f1b,slimpipe,mepipe}. Existing SPP methods target softmax attention and its prefix key/value dependencies. StateFlow extends SPP to recurrent and hybrid models through state-aware chunk scheduling, low-latency state transition, and hybrid-aware partitioning, reducing activation memory and improving training throughput. Table~\ref{tab:related_parallel} compares these methods, with green highlighting favorable entries. It is worth highlighting that SPP can be used in combination with many other parallelism strategies such as DP, TP, and SP.

\section{Method}

Section~\ref{sec:seq-chunk-pipeline} presents the design of StateFlow, an SPP approach for linear recurrence.
% We first describe the chunk-level execution model for linear recurrence.
Then, in Section~\ref{sec:hybrid-chunking}, we discuss sequence partitioning for hybrid models, where recurrent and causal layers favor different partitioning strategies. Section~\ref{sec:stateflow-efficiency-analysis} provides an analysis on how StateFlow reduces memory costs and improves training throughput. Finally, Section~\ref{sec:sandwich-overlap} presents a resource-aware execution strategy to hide the exposed latency of recurrent state transition kernels.

\subsection{Sequence Pipeline Parallelism for Linear Recurrence}
\label{sec:seq-chunk-pipeline}

% \textcolor{orange}{Maybe we should give a brief summary of the method?}

% Figure~\ref{fig:stateflow_seq1f1b} gives an overview by contrasting
% conventional 1F1B pipeline parallelism with StateFlow.
% Figure~\ref{fig:stateflow_seq1f1b}(a) shows conventional 1F1B, where each
% pipeline scheduling unit is the forward or backward execution of one
% full-sequence micro-batch. 
% Figure~\ref{fig:stateflow_seq1f1b}(b) shows StateFlow, which partitions the
% sequence of each micro-batch into coarse-grained outer chunks and exposes the
% forward and backward execution of each outer chunk as a pipeline scheduling
% unit.

% The shading gradient indicates the position of an outer chunk along the
% sequence. During the forward pass, boundary states flow from earlier chunks to
% later chunks, progressively enabling their execution. During the backward
% pass, boundary-state gradients flow in the reverse sequence order. After an outer chunk completes backward execution, StateFlow releases its
% saved tensors and propagates  boundary gradient to the
% preceding chunk.
% The lower part of
% Figure~\ref{fig:stateflow_seq1f1b}(b) illustrates state flow at two levels.
% Across outer chunks, boundary states flow forward and boundary-state gradients
% flow backward. Within each outer chunk, the linear recurrence kernels operate over kernel
% level inner chunks \(b_{i,j}\). Dense tensor kernels expose parallelism across
% these inner chunks, while recurrent state transitions preserve the sequential
% state passing dependency between neighboring inner chunks.

\begin{figure*}[t]
    \centering
    \includegraphics[width=\textwidth]{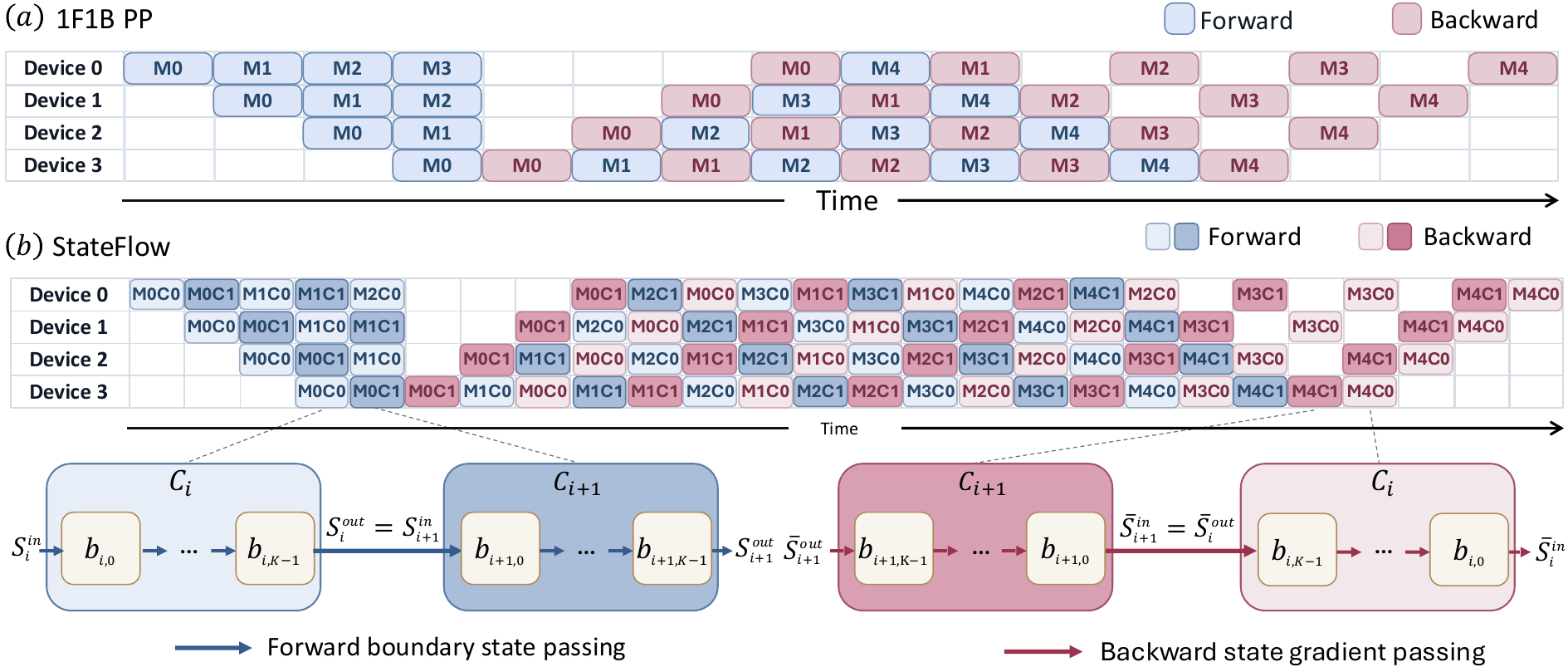}
    \caption{
    Illustration of StateFlow~(ours) and conventional 1F1B, with four devices. (a)~Conventional 1F1B PP schedule where a batch is split into 5 microbatches \(\{\mathrm{M}0,\mathrm{M}1,\cdots\}\), which are used as pipeline units. (b)~StateFlow further partitions each microbatch \(\mathrm{M}a\) into chunks \(\{\mathrm{M}a\mathrm{C}0,\mathrm{M}a\mathrm{C}1,\cdots\}\), where \(\mathrm{M}a\mathrm{C}i\) denotes the \(i\)-th chunk of \(a\)-th microbatch.
    % Blue and red boxes denote forward and backward execution, respectively, while color intensity indicates the chunk position along the sequence.
    % The lower schematics show boundary states and their gradients propagating across outer chunks, and linear recurrence kernels operating over kernel level inner chunks \(b_{i,j}\).
    }
    \label{fig:stateflow_seq1f1b}
\end{figure*}

StateFlow exposes sequence chunks as pipeline scheduling units while preserving
their recurrent dependencies. We first formalize the chunk-level execution
semantics. We consider a single training sequence and omit the microbatch index
for clarity. Let \(S_i^{\mathrm{in}}\) and \(S_i^{\mathrm{out}}\) denote the recurrent
states at the input and output boundaries of chunk \(C_i\), respectively.%
\footnote{In practice, \(S_i^{\mathrm{in}}\) and \(S_i^{\mathrm{out}}\) may
also include auxiliary boundary states, such as short convolution states
commonly used in linear recurrence models.}
% We first define the chunk semantics for a single training sequence and omit the
% micro-batch index for clarity. We consider a linear attention layer
% that maintains a boundary state along the sequence. For notation simplicity, we
% use \(S_t\) to denote the complete state required to continue the computation
% after position \(t\). In standard linear attention, \(S_t\) is the
% matrix-valued recurrent state. In implementations with additional finite-context
% operators, such as short convolution, \(S_t\) can be extended to include the
% corresponding boundary information, such as the convolution tail. Given an
% input sequence \(X=(x_0,x_1,\ldots,x_{T-1})\), the layer updates its state and
% produces a token-level output as
% \begin{equation}
%     S_t = U(S_{t-1}, x_t), \qquad
%     y_t = R(S_t, x_t),
% \end{equation}
% where \(U\) denotes the state transition rule and
% \(R\) denotes the readout computation. This abstraction
% covers different linear attention variants. StateFlow does not depend
% on the specific state transition rule, as long as the boundary state is
% sufficient for continuing the computation of the next sequence segment.
StateFlow partitions a sequence $X$ into chunks:
\begin{equation}
    X = C_0 \Vert C_1 \Vert \cdots \Vert C_{N-1},
\end{equation}
where \(N\) is the number of chunks, \(C_i\) is the \(i\)-th chunk and \(\Vert\) denotes concatenation along the sequence dimension.
% These outer chunks are the units visible to the pipeline scheduler.

\paragraph{Forward Computation}

The forward computation $\mathrm{Fwd}$ of \(C_i\) consumes its input activation \(C^{\mathrm{in}}_{i}\) and input boundary state \(S^{\mathrm{in}}_i\), and produces the chunk output \(C^\mathrm{out}_{i}\) and output boundary state \(S^\mathrm{out}_i\):
\begin{equation}
    \left(
        C_i^\mathrm{out},
        S_i^\mathrm{out}
    \right)
    =
    \mathrm{Fwd}
    \left(
        C_i^\mathrm{in},
        S_i^\mathrm{in}
    \right),
    \qquad 0 \le i < N .
\end{equation}
% Here, \(C_i^\mathrm{in}\) and \(C_i^\mathrm{out}\) denote the input and output activations associated with \(C_i\), and \(S_i^\mathrm{in}\) and \(S_i^\mathrm{out}\) are its input and output boundary states.
The first chunk receives the initial state $S_\mathrm{init}$, while the output state of each chunk becomes the input state of its successor:
\begin{equation}
    S_0^\mathrm{in}=S_{\mathrm{init}},
    \qquad
    S_{i+1}^\mathrm{in}=S_i^\mathrm{out},
    \quad 0 \le i < N-1 .
\end{equation}

\paragraph{Backward Computation}

During the backward pass $\mathrm{Bwd}$, we must preserve the reverse state dependency between chunks.
Let \(\bar{S}_i^\mathrm{out}\) denote the gradient received at the output boundary of chunk \(C_i\). The backward computation of \(C_i\) produces gradients with respect to its input state:
\begin{equation}
\begin{aligned}
    \left(
        \bar{C}_i^{in},
        \bar{S}_i^{in}
    \right)
    =
    \mathrm{Bwd}
    \left(
        C_i^\mathrm{in},
        S_i^\mathrm{in},
        \bar{C}_i^\mathrm{out},
        \bar{S}_i^\mathrm{out}
    \right),
    \quad 0 \le i < N .
\end{aligned}
\end{equation}
Because \(S_{i+1}^\mathrm{in}=S_i^\mathrm{out}\), the input boundary-state gradient
produced by \(C_{i+1}\) becomes the output boundary-state gradient consumed
by \(C_i\):
\begin{equation}
    \bar{S}_i^\mathrm{out}
    =
    \bar{S}_{i+1}^\mathrm{in},
    \qquad 0 \le i < N-1 .
\end{equation}

\paragraph{Scheduling Dependencies}

Formally, let \(P\) denote the number of pipeline stages, and let
\(F_{r,a,i}\) and \(B_{r,a,i}\) denote the forward and backward execution,
respectively, of chunk \(C_i\) from microbatch \(a\) on pipeline stage
\(r\). A legal StateFlow schedule must satisfy the following partial order:
\begin{equation}
\begin{aligned}
    F_{r-1,a,i} &\prec F_{r,a,i},
    && 1 \le r < P, \\
    F_{r,a,i-1} &\prec F_{r,a,i},
    && 1 \le i < N, \\
    F_{r,a,i} &\prec B_{r,a,i},
    && 0 \le r < P,\ 0 \le i < N, \\
    B_{r+1,a,i} &\prec B_{r,a,i},
    && 0 \le r < P-1, \\
    B_{r,a,i+1} &\prec B_{r,a,i},
    && 0 \le i < N-1 .
\end{aligned}
\end{equation}
The first, third, and fourth constraints are standard pipeline dependencies.
The second and fifth constraints arise from forward boundary-state propagation
and reverse propagation of boundary-state gradients between adjacent chunks.
These constraints define schedule correctness but do not prescribe a unique
execution order. StateFlow can issue the ready tasks with different pipeline issue
policies. 

Figure~\ref{fig:stateflow_seq1f1b} contrasts conventional 1F1B with one legal 1F1B-style StateFlow schedule. Figure~\ref{fig:stateflow_seq1f1b}~(a) shows conventional 1F1B, where each pipeline scheduling unit is the forward or backward pass of one full-sequence microbatch \(\{\mathrm{M}0,\mathrm{M}1,\cdots\}\). Figure~\ref{fig:stateflow_seq1f1b}~(b) shows StateFlow, which partitions the sequence of each microbatch \(\mathrm{M}a\) into chunks \(\{\mathrm{M}a\mathrm{C}0,\mathrm{M}a\mathrm{C}1,\cdots\}\) and exposes the forward and backward pass of each chunk as a pipeline scheduling unit. 
StateFlow issues ready tasks while respecting pipeline device and recurrent state dependencies.

% The shading gradient indicates the position of each chunk along the sequence.
During forward execution, boundary-states propagate from earlier chunks to later chunks. During backward execution, boundary-state gradients propagate in the reverse order, and the tensors saved for a chunk can be released after its backward execution completes. The bottom part in Figure~\ref{fig:stateflow_seq1f1b}~(b) shows the two chunk levels.  Within each chunk, the existing linear recurrence kernels operate over kernel-level chunks \(b_{i,j}\). Dense tensor kernels expose parallelism across these kernel-level chunks, while recurrent state transitions preserve the sequential dependency between neighboring kernel-level chunks. 

StateFlow treats the number of sequence chunks \(N\) as a tunable scheduling parameter and selects the fastest feasible value from \(N\in\{2,4,6,8,16,32\}\) within the GPU memory constraint.

\subsection{Hybrid-Aware Sequence Chunking}
\label{sec:hybrid-chunking}

To keep the pipeline bubbles small, the runtime of chunks should be balanced. Equal-length chunks are effective for linear recurrence because their FLOPs depend only on the chunk size. However, for hybrid models, later chunks perform more softmax attention work because they attend to longer prefixes. 

Let a sequence of length \(T\) be partitioned into \(N\) chunks with boundaries \(0=c_0<c_1<\cdots<c_N=T\), where \(s_i=c_{i+1}-c_i\) is the length of the $i$-th chunk. The computation cost of the \(i\)-th chunk can be formulated as
\begin{equation}
\mathrm{Cost}_i
=
\left(
    L_{\mathrm{lin}}C_{\mathrm{lin}}
    + C_{\mathrm{dense}}
\right)s_i
+
L_{\mathrm{soft}}C_{\mathrm{soft}}
\left(
    c_i s_i+\frac{s_i^2}{2}
\right).
\label{eq:hybrid_chunk_cost}
\end{equation}
Here, \(L_{\mathrm{lin}}\) and \(L_{\mathrm{soft}}\) are the numbers of linear recurrence and softmax attention layers. \(C_{\mathrm{lin}}\) is the per-token FLOPs of one linear recurrence layer, \(C_{\mathrm{soft}}\) is the FLOP coefficient per query--key pair for one softmax attention layer, and \(C_{\mathrm{dense}}\) represents the remaining token-wise dense FLOPs. The FLOP-balanced partition satisfies
\begin{equation}
    \mathrm{Cost}_0 = \mathrm{Cost}_1 = \cdots = \mathrm{Cost}_{N-1}.
    \label{eq:hybrid_boundary_condition}
\end{equation}

% For brevity, let \(W_{\mathrm{tok}}= L_{\mathrm{lin}}C_{\mathrm{lin}}+C_{\mathrm{dense}}\) and \(W_{\mathrm{soft}}= L_{\mathrm{soft}}C_{\mathrm{soft}}\). The accumulated proxy cost over the prefix interval \([0,c)\) is
% \begin{equation}
% \mathrm{Cost}_{\mathrm{acc}}(c)
% =
% W_{\mathrm{tok}}c
% +
% \frac{W_{\mathrm{soft}}}{2}c^2 .
% \label{eq:hybrid_accumulated_cost}
% \end{equation}
% StateFlow obtains the hybrid balanced boundaries by equalizing this accumulated
% cost:
% \begin{equation}
% \mathrm{Cost}_{\mathrm{acc}}
% \left(c_i^{\mathrm{hybrid}}\right)
% =
% \frac{i}{N}
% \mathrm{Cost}_{\mathrm{acc}}(T),
% \qquad i=0,\ldots,N .
% \label{eq:hybrid_boundary_condition}
% \end{equation}
% When \(W_{\mathrm{soft}}=0\), this condition naturally reduces to the
% equal-length partition \(c_i^{\mathrm{equal}}=iT/N\).

\paragraph{Optimal Chunk Search}

FLOPs-based balancing does not capture kernel efficiency and memory behavior, etc. StateFlow therefore uses the hybrid balanced partition as an anchor and generates candidates
\begin{equation}
c_i(\alpha)
=
(1-\alpha)c_i^{\mathrm{equal}}
+
\alpha c_i^{\mathrm{hybrid}},
\qquad i=0,\ldots,N,
\label{eq:hybrid_partition_interpolation}
\end{equation} 
where \(c_i^{\mathrm{equal}}=iT/N\) and $c_i^\mathrm{hybrid}$ is the chunk boundary that satisfies Eq~\ref{eq:hybrid_boundary_condition}. When \(\alpha=0\), we get equal-length chunking, while \(\alpha=1\) gives the FLOP-balanced partition, and \(\alpha>1\) explores more aggressive imbalance. The generated boundaries are rounded to the closest multiple of 128.

For hybrid models, StateFlow profiles \(N\in\{2,4,6,8,16,32\}\) and \(\alpha\in\{0,0.1,0.2,\ldots,1.0,1.25\}\), and selects the fastest feasible pair within the GPU memory constraint.

\begin{figure*}[t]
    \centering
    \includegraphics[width=\textwidth]
    {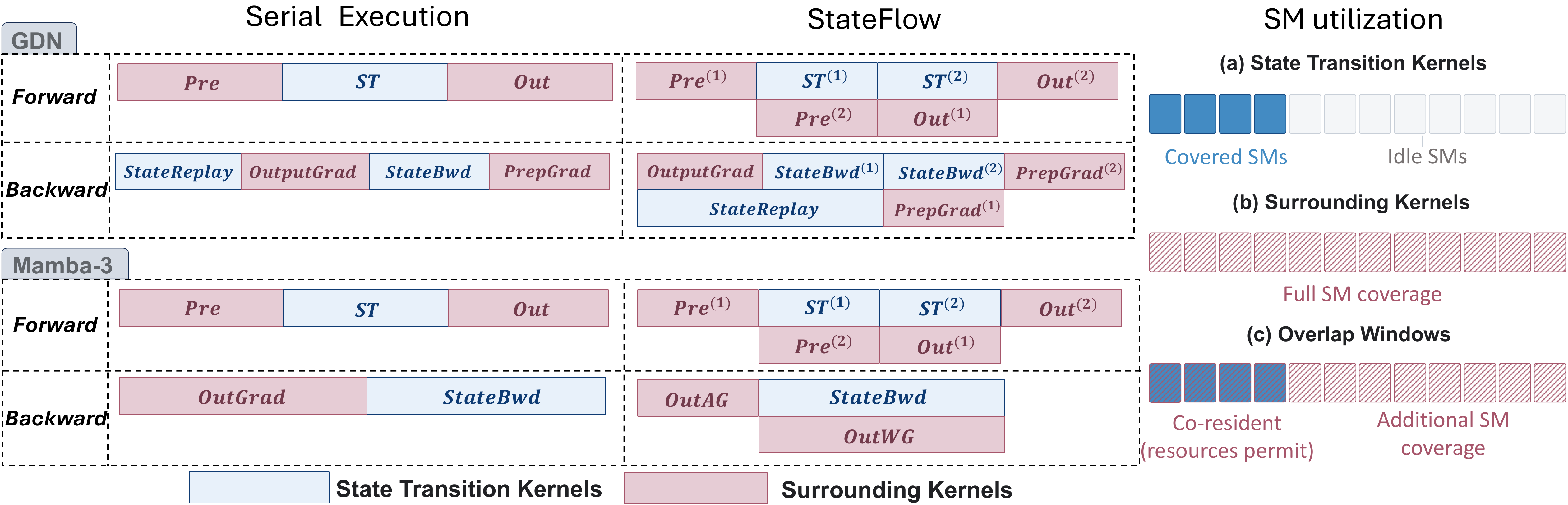}
    \caption{
    Forward and backward overlap schedules for GDN and Mamba-3.
    The SM utilization diagrams show how surrounding kernels use idle SMs and
    may co-reside with state transition kernels when resources permit.
    }
    \Description{
    Serial and StateFlow execution for GDN and Mamba-3.
    StateFlow overlaps state transition kernels with surrounding forward and
    backward computation. The right side illustrates SM coverage during serial
    and concurrent execution.
    }
    \label{fig:sandwich-overlap}
\end{figure*}

\subsection{Efficiency Analysis}
\label{sec:stateflow-efficiency-analysis}

We analyze how StateFlow affects peak memory and training throughput for linear recurrent models under the 1F1B-style schedule shown in Figure~\ref{fig:stateflow_seq1f1b}. Conventional 1F1B schedules \(M\) full-sequence microbatches as pipeline units, whereas StateFlow partitions each microbatch into \(N\) chunks, yielding \(MN\) chunk-level scheduling units.

\paragraph{Memory Reduction}

Let \(A_r\) denote the activation footprint of one full-sequence microbatch
on device \(r\), and let \(B_r\) denote the boundary-information footprint
of one chunk on device \(r\). In conventional 1F1B, device \(r\) stores at most \(\min(M,P-r)\) microbatches:
\begin{equation}
    \mathrm{Mem}_{\mathrm{1F1B}}(r)
    \approx
    \min(M,P-r)A_r .
    \label{eq:mem_mb}
\end{equation}
StateFlow must store one complete \(N\)-chunk sequence before backward, together with up to \(P-r-1\) additional chunks caused by the downstream pipeline delay. Its memory cost is therefore approximated by
\begin{equation}
    \mathrm{Mem}_{\mathrm{SF}}(r)
    \approx
    \min\left(MN,N+P-r-1\right)
    \left(
        \frac{A_r}{N}+B_r
    \right).
    \label{eq:mem_stateflow}
\end{equation}

For linear recurrence models, \(B_r\) consists of compact recurrent and
local-convolution boundary states and usually satisfies
\(B_r\ll A_r/N\). When \(M\ge P-r\) and \(MN\ge N+P-r-1\), StateFlow reduces the retained activation from \(A_r(P-r)\) to \(A_r\left(1+(P-r-1)/N\right)\). 

For hybrid models, \(B_r\) additionally accounts for the K/V tensors retained
by softmax attention layers and may therefore be non-negligible. To avoid
duplicating prefix K/V across chunks, StateFlow stores each microbatch's K/V
tensors only once in sequence-wide buffers. These buffers are also retained
under activation recomputation to support efficient chunk replay. This design
substantially reduces the remaining boundary overhead, so the approximately
\(1/N\) reduction in activation footprint generally outweighs this additional
cost, allowing StateFlow to reduce peak memory in both layer types.

% The memory savings are larger on earlier pipeline stages, where conventional
% 1F1B retains more in-flight microbatches, thereby reducing both the global
% peak memory and the memory imbalance across stages.

Hybrid-aware partitioning might slightly increase peak activation memory.
Forward dependencies execute chunks in the sequence order, whereas backward execution releases their saved activations in reverse. Thus, later chunks are released first, while earlier chunks need to remain in memory longer. StateFlow bounds the resulting memory overhead by treating the GPU memory limit as a hard constraint during candidate profiling.

\paragraph{Throughput Gains}

The same decrease in scheduling granularity also reduces pipeline bubbles, leading to higher throughput. Let
\(t\) denote the critical-path forward-backward time of one full-sequence
microbatch. Conventional 1F1B takes approximately
\begin{equation}
    T_{\mathrm{1F1B}}
    \approx
    (M+P-1)t .
    \label{eq:time_mb}
\end{equation}
If sequence chunks scale ideally, each chunk takes roughly \(t/N\), and
StateFlow takes
\begin{equation}
    T_{\mathrm{SF}}
    \approx
    (MN+P-1)\frac{t}{N}
    =
    \left(
        M+\frac{P-1}{N}
    \right)t .
    \label{eq:time_stateflow}
\end{equation}
This comparison shows that exposing sequence chunks as scheduling units
amortizes pipeline bubbles across \(N\) chunks, leading to better hardware
utilization, while overly fine chunks may reduce kernel efficiency.

\subsection{State Transition Grid Optimization and Overlap}
\label{sec:sandwich-overlap}

In StateFlow, the microbatch size of each scheduling unit is very small~(typically set to 1). In such cases, the state transition~(ST) kernel has limited streaming multiprocessor~(SM) coverage, leading to hardware underutilization.
% StateFlow's chunk schedule keeps only one active microbatch within each sequence chunk, so the state transition kernel often covers only a subset of SMs with limited occupancy. Because the next outer chunk must wait for its boundary state, this underutilized execution remains exposed on the pipeline critical path. 
StateFlow mitigates this with~(1) state transition grid optimization and~(2) split and overlap.

\subsubsection{State Transition Grid Optimization}

For state transition implementations that are tiled along the value dimension, the launch grid size of the ST kernel is roughly
\begin{equation}
\label{eq:state-launch-grid}
    G
    =
    B_{\mathrm{chunk}} A
    \left\lceil \frac{d_v}{B_V} \right\rceil ,
\end{equation}
where \(B_{\mathrm{chunk}}\) is the active microbatch count within a chunk and is set to one in the memory-minimal schedule. \(A\) is the number of linear recurrence heads and is fixed by the model and TP configuration, \(d_v\) is the value dimension, and \(B_V\) is the tile size along that dimension.

The value dimension tiling therefore becomes the main tuning knob. Reducing \(B_V\) increases the number of launched blocks and improves SM coverage, but an overly small tile reduces kernel efficiency. StateFlow jointly profiles \(B_V\in\{8,16,32,64\}\) and selects the configuration
with the lowest measured latency.

\subsubsection{Split and Overlap}

\paragraph{Forward Pass}
As mentioned in Section~\ref{sec:chunk-wise-parallelism}, each chunk with explicit recurrence kernels can be split into three stages: preparation \(Pre\), state transition \(ST\), and output computation \(Out\). For each token/chunk, these stages must be computed sequentially, but stages from different tokens/chunk can be computed in parallel to maximize hardware utilization.

As shown in Figure~\ref{fig:sandwich-overlap}, StateFlow splits each chunk along the sequence dimension into two contiguous segments. It overlaps the state transition of the first segment, \(ST^{(1)}\), with the preparation of the second segment, \(Pre^{(2)}\), and then overlaps the state transition of the second segment, \(ST^{(2)}\), with the output computation of the first segment, \(Out^{(1)}\):
\begin{equation}
\label{eq:fwd-split-schedule}
    Pre^{(1)}
    \rightarrow
    \left(
        {ST}^{(1)} \parallel Pre^{(2)}
    \right)
    \rightarrow
    \left(
        {ST}^{(2)} \parallel Out^{(1)}
    \right)
    \rightarrow
    Out^{(2)} .
\end{equation}

The same forward wavefront also applies to the fused Mamba-3 kernel, where
\(ST\) denotes the fused state transition, \(Pre\) groups gating and projection
computations preceding \(ST\), and \(Out\) groups output and MLP projections
following \(ST\) within the same chunk.

\definecolor{sfblue}{HTML}{EEF4FF}
\definecolor{modelgray}{HTML}{F4F4F4}

\begin{table*}[!t]
    \centering
    \caption{
        End-to-end training efficiency results with activation recomputation for (a) GDN and (b) Mamba-3. Each entry reports throughput/peak memory in K tokens/s and GB. Megatron and Swift denote their native pipeline baselines, while ``+StateFlow'' uses the highest-throughput feasible chunk count \(N\) for each workload. OOM denotes out of memory.
    }
    \label{tab:exp1_full_recompute_all}

    \fontsize{9.1pt}{9.1pt}\selectfont

    \centerline{\textbf{(a) GDN}}
\begin{tblr}{
        width = \textwidth,
        colspec = {
            Q[c,m,wd=0.095\textwidth]
            *{9}{Q[c,m,wd=0.097\textwidth]}
        },
        rowsep = 0.8pt,
        colsep = 0.16pt,
        cells = {c,m},
        column{1} = {c,m},
        hline{1,2,3,7,8,12,13,Z} = {0.45pt, solid},
        vline{2,5,8} = {0.35pt, solid},
        row{1} = {
            font=\bfseries,
            abovesep=1.0pt,
            belowsep=1.0pt
        }
    }

    Method
    & GBS4 & GBS8 & GBS16
    & GBS4 & GBS8 & GBS16
    & GBS4 & GBS8 & GBS16 \\

    \SetRow{
        bg=modelgray,
        font=\bfseries,
        abovesep=0.8pt,
        belowsep=0.8pt
    }
    \SetCell{c,m} 3B
    & \SetCell[c=3]{c,m} 64K
    & &
    & \SetCell[c=3]{c,m} 128K
    & &
    & \SetCell[c=3]{c,m} 256K
    & & \\

    \SetRow{abovesep=1.2pt, belowsep=1.2pt}
    \SetCell{c,m} Megatron
    & 20.2/43.7G
    & 28.7/43.7G
    & 35.4/43.7G
    & 18.8/75.3G
    & 28.4/76.5G
    & 35.7/76.5G
    & OOM
    & OOM
    & OOM \\

    \SetRow{
        bg=sfblue,
        abovesep=1.2pt,
        belowsep=1.2pt
    }
    \SetCell{c,m,bg=sfblue}
    +StateFlow
    & \textbf{38.6}/24.3G
    & \textbf{42.3}/26.3G
    & \textbf{43.2}/27.5G
    & \textbf{41.4}/40.6G
    & \textbf{43.8}/41.3G
    & \textbf{44.2}/46.5G
    & \textbf{43.7}/69.9G
    & \textbf{45.0}/74.8G
    & \textbf{45.7}/73.4G \\

    \SetRow{abovesep=1.2pt, belowsep=1.2pt}
    \SetCell{c,m} Swift
    & 19.3/25.5G
    & 27.4/26.6G
    & 33.9/26.7G
    & 19.3/46.5G
    & 27.3/48.8G
    & 32.3/48.8G
    & OOM
    & OOM
    & OOM \\

    \SetRow{
        bg=sfblue,
        abovesep=1.2pt,
        belowsep=1.2pt
    }
    \SetCell{c,m,bg=sfblue}
    +StateFlow
    & \textbf{38.3}/19.5G
    & \textbf{40.6}/19.7G
    & \textbf{41.7}/19.7G
    & \textbf{40.5}/34.3G
    & \textbf{42.2}/34.3G
    & \textbf{43.1}/34.3G
    & \textbf{42.1}/64.3G
    & \textbf{43.1}/64.0G
    & \textbf{44.0}/64.0G \\

    \SetRow{
        bg=modelgray,
        font=\bfseries,
        abovesep=0.8pt,
        belowsep=0.8pt
    }
    \SetCell{c,m} 15B
    & \SetCell[c=3]{c,m} 64K
    & &
    & \SetCell[c=3]{c,m} 128K
    & &
    & \SetCell[c=3]{c,m} 256K
    & & \\

    \SetRow{abovesep=1.2pt, belowsep=1.2pt}
    \SetCell{c,m} Megatron
    & 12.7/49.8G
    & 15.9/49.8G
    & 18.2/49.8G
    & OOM
    & OOM
    & OOM
    & OOM
    & OOM
    & OOM \\

    \SetRow{
        bg=sfblue,
        abovesep=1.2pt,
        belowsep=1.2pt
    }
    \SetCell{c,m,bg=sfblue}
    +StateFlow
    & \textbf{16.9}/29.8G
    & \textbf{18.3}/29.8G
    & \textbf{19.2}/35.7G
    & \textbf{18.3}/41.4G
    & \textbf{19.1}/44.3G
    & \textbf{20.0}/44.5G
    & \textbf{19.1}/65.3G
    & \textbf{19.9}/66.8G
    & \textbf{20.5}/72.2G \\

    \SetRow{abovesep=1.2pt, belowsep=1.2pt}
    \SetCell{c,m} Swift
    & 13.4/40.5G
    & 17.0/40.5G
    & 19.5/40.6G
    & OOM
    & OOM
    & OOM
    & OOM
    & OOM
    & OOM \\

    \SetRow{
        bg=sfblue,
        abovesep=1.2pt,
        belowsep=1.2pt
    }
    \SetCell{c,m,bg=sfblue}
    +StateFlow
    & \textbf{19.9}/20.1G
    & \textbf{20.9}/20.1G
    & \textbf{21.3}/21.4G
    & \textbf{20.9}/31.1G
    & \textbf{21.3}/31.5G
    & \textbf{21.9}/31.5G
    & \textbf{21.3}/53.5G
    & \textbf{21.9}/53.5G
    & \textbf{22.2}/54.4G \\

    \SetRow{
        bg=modelgray,
        font=\bfseries,
        abovesep=0.8pt,
        belowsep=0.8pt
    }
    \SetCell{c,m} 32B
    & \SetCell[c=3]{c,m} 32K
    & &
    & \SetCell[c=3]{c,m} 64K
    & &
    & \SetCell[c=3]{c,m} 128K
    & & \\

    \SetRow{abovesep=1.2pt, belowsep=1.2pt}
    \SetCell{c,m} Megatron
    & 8.4/45.7G
    & 10.6/46.2G
    & 12.2/46.2G
    & 8.7/74.2G
    & 10.9/75.3G
    & 12.6/75.3G
    & OOM
    & OOM
    & OOM \\

    \SetRow{
        bg=sfblue,
        abovesep=1.2pt,
        belowsep=1.2pt
    }
    \SetCell{c,m,bg=sfblue}
    +StateFlow
    & \textbf{9.9}/34.7G
    & \textbf{11.5}/34.7G
    & \textbf{12.4}/34.7G
    & \textbf{11.4}/40.7G
    & \textbf{12.4}/40.7G
    & \textbf{13.2}/52.3G
    & \textbf{12.4}/52.8G
    & \textbf{13.2}/64.3G
    & \textbf{13.8}/64.3G \\

    \SetRow{abovesep=1.2pt, belowsep=1.2pt}
    \SetCell{c,m} Swift
    & 9.3/38.5G
    & 11.8/39.6G
    & 13.7/39.6G
    & 9.5/67.3G
    & 12.0/68.5G
    & 13.8/70.5G
    & OOM
    & OOM
    & OOM \\

    \SetRow{
        bg=sfblue,
        abovesep=1.2pt,
        belowsep=1.2pt
    }
    \SetCell{c,m,bg=sfblue}
    +StateFlow
    & \textbf{11.4}/27.2G
    & \textbf{13.5}/21.4G
    & \textbf{14.4}/27.2G
    & \textbf{13.6}/27.5G
    & \textbf{14.5}/33.1G
    & \textbf{15.2}/33.2G
    & \textbf{14.5}/45.5G
    & \textbf{15.2}/45.5G
    & \textbf{15.5}/57.1G \\

    \end{tblr}

    \smallskip
    \textbf{(b) Mamba-3}\par
\begin{tblr}{
        width = \textwidth,
        colspec = {
            Q[c,m,wd=0.095\textwidth]
            *{9}{Q[c,m,wd=0.097\textwidth]}
        },
        rowsep = 0.8pt,
        colsep = 0.16pt,
        cells = {c,m},
        column{1} = {c,m},
        hline{1,2,3,7,8,12,13,Z} = {0.45pt, solid},
        vline{2,5,8} = {0.35pt, solid},
        row{1} = {
            font=\bfseries,
            abovesep=1.0pt,
            belowsep=1.0pt
        }
    }

    Method
    & GBS4 & GBS8 & GBS16
    & GBS4 & GBS8 & GBS16
    & GBS4 & GBS8 & GBS16 \\

    \SetRow{
        bg=modelgray,
        font=\bfseries,
        abovesep=0.8pt,
        belowsep=0.8pt
    }
    \SetCell{c,m} 3B
    & \SetCell[c=3]{c,m} 64K
    & &
    & \SetCell[c=3]{c,m} 128K
    & &
    & \SetCell[c=3]{c,m} 256K
    & & \\

    \SetRow{abovesep=1.2pt, belowsep=1.2pt}
    \SetCell{c,m} Megatron
    & 16.1/43.0G
    & 22.8/43.0G
    & 28.8/43.0G
    & OOM
    & OOM
    & OOM
    & OOM
    & OOM
    & OOM \\

    \SetRow{
        bg=sfblue,
        abovesep=1.2pt,
        belowsep=1.2pt
    }
    \SetCell{c,m,bg=sfblue}
    +StateFlow
    & \textbf{33.7}/25.7G
    & \textbf{35.3}/28.7G
    & \textbf{36.8}/29.1G
    & \textbf{35.4}/42.5G
    & \textbf{36.9}/42.5G
    & \textbf{37.7}/42.5G
    & \textbf{36.9}/70.5G
    & \textbf{37.8}/70.4G
    & \textbf{38.2}/70.6G \\

    \SetRow{abovesep=1.2pt, belowsep=1.2pt}
    \SetCell{c,m} Swift
    & 15.6/26.0G
    & 22.1/31.0G
    & 27.9/31.0G
    & 15.6/47.3G
    & 22.0/57.4G
    & 27.7/57.4G
    & OOM
    & OOM
    & OOM \\

    \SetRow{
        bg=sfblue,
        abovesep=1.2pt,
        belowsep=1.2pt
    }
    \SetCell{c,m,bg=sfblue}
    +StateFlow
    & \textbf{32.3}/19.8G
    & \textbf{33.9}/20.1G
    & \textbf{35.4}/20.1G
    & \textbf{33.9}/34.7G
    & \textbf{35.4}/34.7G
    & \textbf{36.2}/35.3G
    & \textbf{35.2}/64.7G
    & \textbf{36.0}/64.4G
    & \textbf{36.6}/64.5G \\

    \SetRow{
        bg=modelgray,
        font=\bfseries,
        abovesep=0.8pt,
        belowsep=0.8pt
    }
    \SetCell{c,m} 15B
    & \SetCell[c=3]{c,m} 64K
    & &
    & \SetCell[c=3]{c,m} 128K
    & &
    & \SetCell[c=3]{c,m} 256K
    & & \\

    \SetRow{abovesep=1.2pt, belowsep=1.2pt}
    \SetCell{c,m} Megatron
    & 9.0/52.9G
    & 11.4/53.0G
    & 13.2/53.0G
    & OOM
    & OOM
    & OOM
    & OOM
    & OOM
    & OOM \\

    \SetRow{
        bg=sfblue,
        abovesep=1.2pt,
        belowsep=1.2pt
    }
    \SetCell{c,m,bg=sfblue}
    +StateFlow
    & \textbf{11.9}/31.2G
    & \textbf{12.9}/31.6G
    & \textbf{13.8}/38.6G
    & \textbf{12.8}/42.1G
    & \textbf{13.8}/45.2G
    & \textbf{14.4}/46.2G
    & \textbf{13.8}/68.8G
    & \textbf{14.4}/68.5G
    & \textbf{14.9}/72.9G \\

    \SetRow{abovesep=1.2pt, belowsep=1.2pt}
    \SetCell{c,m} Swift
    & 9.4/32.4G
    & 11.9/32.5G
    & 13.7/32.5G
    & OOM
    & OOM
    & OOM
    & OOM
    & OOM
    & OOM \\

    \SetRow{
        bg=sfblue,
        abovesep=1.2pt,
        belowsep=1.2pt
    }
    \SetCell{c,m,bg=sfblue}
    +StateFlow
    & \textbf{14.2}/21.1G
    & \textbf{14.8}/21.1G
    & \textbf{15.2}/21.9G
    & \textbf{14.7}/34.0G
    & \textbf{15.2}/34.8G
    & \textbf{15.5}/34.8G
    & \textbf{15.1}/57.7G
    & \textbf{15.5}/58.1G
    & \textbf{15.8}/59.1G \\

    \SetRow{
        bg=modelgray,
        font=\bfseries,
        abovesep=0.8pt,
        belowsep=0.8pt
    }
    \SetCell{c,m} 32B
    & \SetCell[c=3]{c,m} 32K
    & &
    & \SetCell[c=3]{c,m} 64K
    & &
    & \SetCell[c=3]{c,m} 128K
    & & \\

    \SetRow{abovesep=1.2pt, belowsep=1.2pt}
    \SetCell{c,m} Megatron
    & 6.0/48.6G
    & 7.5/48.6G
    & 8.7/48.8G
    & OOM
    & OOM
    & OOM
    & OOM
    & OOM
    & OOM \\

    \SetRow{
        bg=sfblue,
        abovesep=1.2pt,
        belowsep=1.2pt
    }
    \SetCell{c,m,bg=sfblue}
    +StateFlow
    & \textbf{8.2}/31.0G
    & \textbf{8.9}/31.0G
    & \textbf{9.3}/31.1G
    & \textbf{8.9}/37.0G
    & \textbf{9.3}/37.1G
    & \textbf{9.7}/43.0G
    & \textbf{9.3}/54.9G
    & \textbf{9.7}/55.0G
    & \textbf{9.9}/55.1G \\

    \SetRow{abovesep=1.2pt, belowsep=1.2pt}
    \SetCell{c,m} Swift
    & 6.3/40.5G
    & 8.0/41.5G
    & 9.2/41.7G
    & OOM
    & OOM
    & OOM
    & OOM
    & OOM
    & OOM \\

    \SetRow{
        bg=sfblue,
        abovesep=1.2pt,
        belowsep=1.2pt
    }
    \SetCell{c,m,bg=sfblue}
    +StateFlow
    & \textbf{8.7}/22.6G
    & \textbf{9.5}/22.7G
    & \textbf{9.9}/22.8G
    & \textbf{9.4}/28.8G
    & \textbf{9.9}/28.9G
    & \textbf{10.3}/34.9G
    & \textbf{9.8}/47.2G
    & \textbf{10.3}/47.1G
    & \textbf{10.5}/47.3G \\

    \end{tblr}
\end{table*}

\paragraph{Backward Pass}
During backward propagation, StateFlow overlaps backward state transition
with state replay and gradient computation.
For GDN, we group the computation into state replay \({SR}\), output
gradient computation \({OG}\), backward state transition
\({STB}\), and preparation gradient computation \({PG}\). As shown in Figure~\ref{fig:sandwich-overlap}, 
StateFlow uses the schedule
\begin{equation}
\label{eq:bwd-overlap-schedule}
    \left[
        {SR}
        \parallel
        \left(
            {OG}
            \rightarrow
            {STB}^{(1)}
        \right)
    \right]
    \rightarrow
    \left(
        {PG}^{(1)}
        \parallel
        {STB}^{(2)}
    \right)
    \rightarrow
    {PG}^{(2)} .
\end{equation}
The first overlap window executes state replay concurrently with output
gradient computation and the first state backward fragment. Once the required
replayed states and state gradients are available, the second state backward
fragment overlaps with dependency-ready post gradient computation.

For Mamba-3, StateFlow decomposes the output-projection backward computation
\(OutPG\) into its activation-gradient \(OutAG\) and weight-gradient \(OutWG\)
components. Since \(OutWG\) is not required by the recurrent gradient
dependency path, StateFlow executes \(OutAG\) first and overlaps the deferred
\(OutWG\) with the fused backward state computation \(StateBwd\):
\begin{equation}
\label{eq:mamba3-bwd-overlap}
    OutAG
    \rightarrow
    \left(
        StateBwd
        \parallel
        OutWG
    \right).
\end{equation}

Figure~\ref{fig:sandwich-overlap} illustrates the overlap opportunity.
Figure~\ref{fig:sandwich-overlap}~(a) shows that state transition kernels
cover only a subset of the SMs, leaving the remaining SMs idle.
Figure~\ref{fig:sandwich-overlap}~(b) shows that the surrounding kernels
launch larger grids and can cover the GPU more fully.
During the overlap in Figure~\ref{fig:sandwich-overlap}~(c), these surrounding
kernels execute on otherwise idle SMs and may also co-reside with the state
transition kernels on active SMs when resources permit, thereby hiding part of
the state transition latency.

For each operator shape, StateFlow profiles a small set of valid tile sizes,
split points, and overlap schedules, and selects the configuration with the
lowest measured latency.

\section{Evaluation}
\label{sec:evaluation}

In this section, we will first describe the evaluation details~(Section~\ref{sec:experimental-setup}). Then, we report the results on linear recurrence models~(Section~\ref{sec:exp-end-to-end}) and hybrid models~(Section~\ref{sec:exp_hybrid}). Finally, we present the efficiency gains on recurrence kernels using split and overlap~(Section~\ref{sec:exp_state_passing_overlap}).

\begin{figure}[!t]
    \centering
    \includegraphics[width=\columnwidth]
    {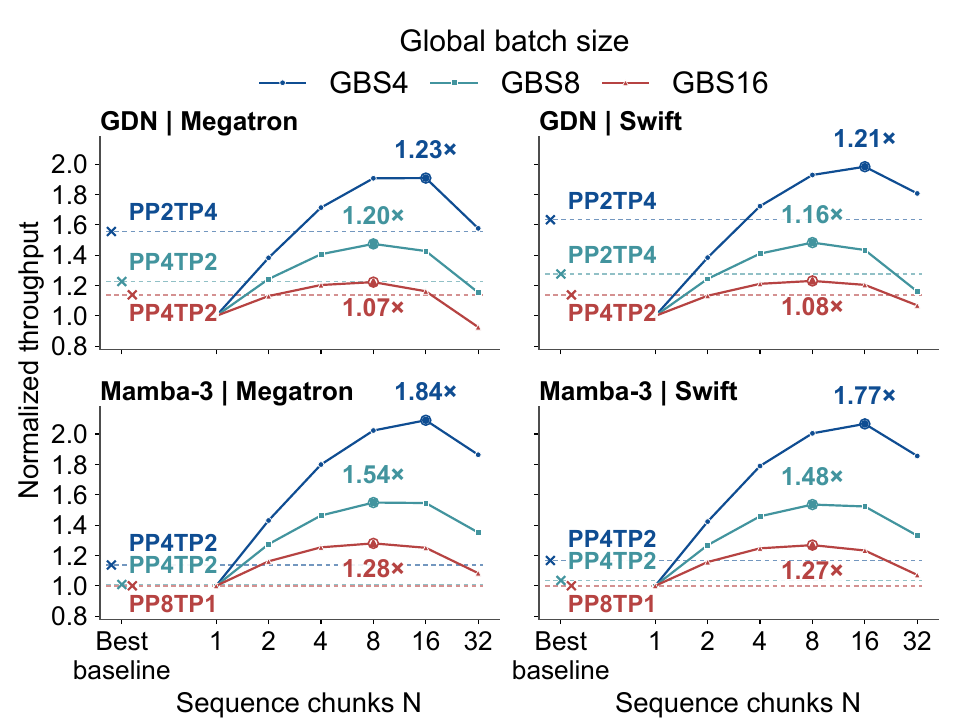}
    \caption{
        Comparison with the best-tuned native baseline and sensitivity to the sequence chunk count \(N\) for 3B linear recurrence models at 64K on eight GPUs. Throughput is normalized to native PP8/TP1. Dashed lines mark the best native configuration, circled points mark the best \(N\), and annotations report speedup over the dashed baseline.
    }
    \Description{
    Normalized throughput versus \(N\) for GDN and Mamba-3 on
    Megatron and Swift at GBS 4, 8, and 16.
    }
    \label{fig:exp1_n_sensitivity}
\end{figure}

\subsection{Experimental Setup}
\label{sec:experimental-setup}

\paragraph{Implementation Details}

Experiments use NVIDIA A100-SXM4-80GB GPUs, with 8, 16, and 32 GPUs for the 3B, 15B, and 32B models, respectively. All experiments use PyTorch 2.6.0,
CUDA 12.4, and Triton 3.5.1.  We implement StateFlow in Megatron Core r0.12.0 and ms-swift 3.6.4~\citep{swift}, using their native pipeline implementations as baselines under identical model, recomputation, and parallelism configurations. Unless otherwise specified, all methods use full activation recomputation in uniform one-layer units. Throughput excludes warmup iterations, and peak memory is the maximum PyTorch CUDA allocation across all GPUs. 
StateFlow is mathematically exact and introduces no approximation.

Unless otherwise specified, we use PP8/TP1, PP4/TP4, and PP4/TP8 for the 3B, 15B, and 32B models, respectively. All pipeline-parallel experiments use
the 1F1B schedule. Each microbatch contains one training
sequence (\(\mathrm{MBS}=1\)). We evaluate
\(\mathrm{GBS}\in\{4,8,16\}\), where
\(\mathrm{GBS}=\mathrm{MBS}\times\mathrm{DP}\times M\) and \(M\) is the
number of microbatches processed per parameter update. Since all configurations use \(\mathrm{DP}=1\), the three GBS values
correspond to \(M=4,8,16\), respectively. StateFlow partitions each microbatch
into \(N\) sequence chunks for scheduling without changing MBS or GBS.

\paragraph{Model Configurations.}
% We experiment with well-established Transformer architecture, but with softmax attention replaced with linear recurrence.
We use the official default hyperparameters for GDN and Mamba-3, with Mamba-3
layers having slightly more parameters. The 3B, 15B, and 32B
models use 32/40/64 layers and hidden sizes 2560/5120/6144. All models use a
50,304-token vocabulary, untied token embeddings and output weights,
pre-LayerNorm blocks, and SwiGLU MLPs. Their GDN and
softmax attention layers use 20/40/64 attention heads, while Mamba-3 uses
80/160/192 SSM heads with head dimension 64 and state dimension 128. Hybrid
models use a 3:1 recurrence-to-attention ratio, with GQA group sizes 4/5/4 for softmax attention~\citep{gqa}.

\begin{figure*}[t]
    \centering
    \includegraphics[width=\textwidth]
    {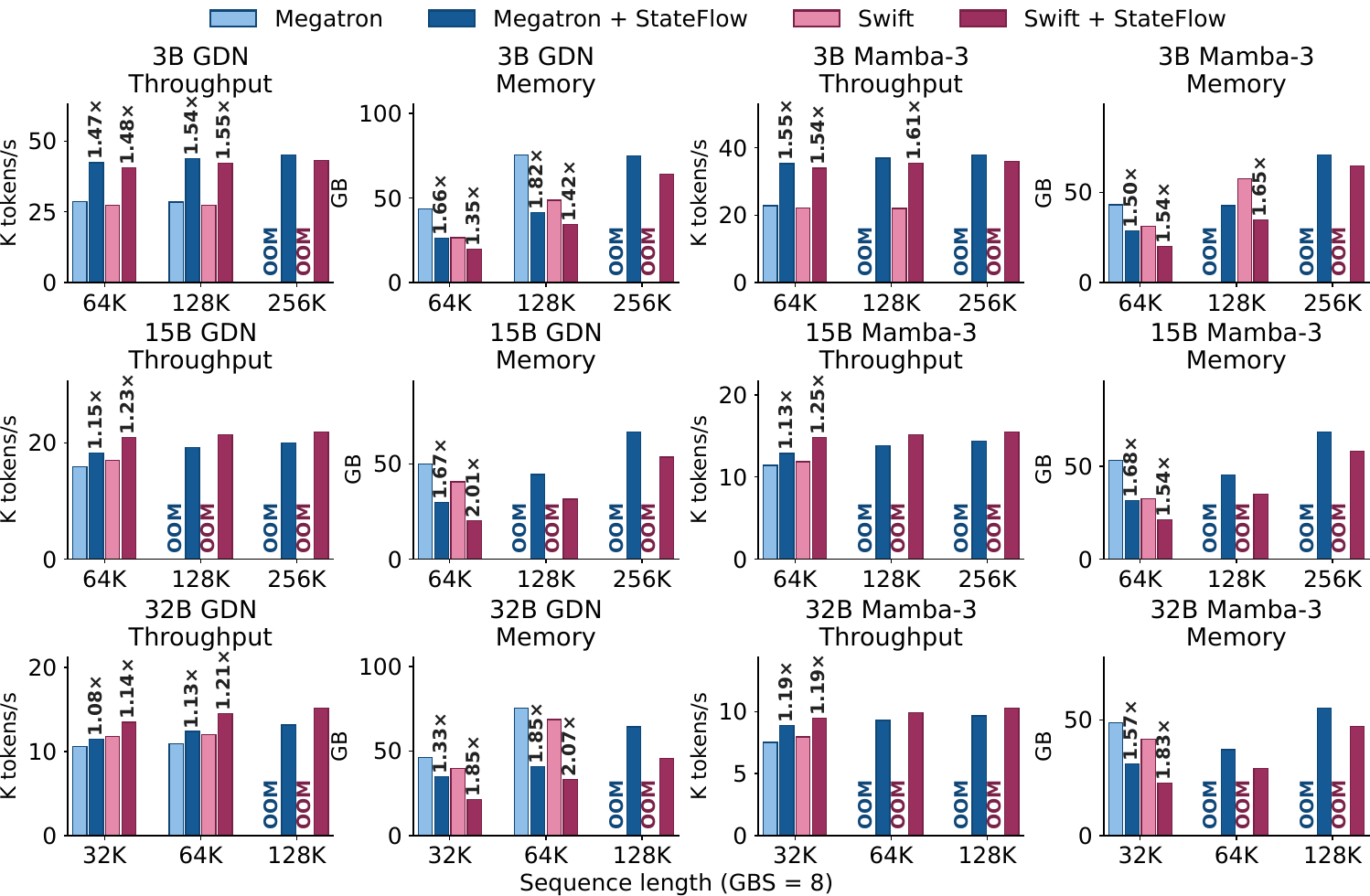}

    \caption{
        End-to-end throughput and peak memory at \(GBS=8\) with activation
        recomputation. Light and dark bars show the native pipelines and StateFlow using the fastest feasible \(N\). Annotations show throughput speedup and peak memory reduction. OOM denotes out of memory.
    }

    \Description{
    Twelve panels compare GDN and Mamba-3 at 3B, 15B, and 32B scales on
    Megatron and Swift across three sequence lengths.
    }
    \label{fig:exp1_gbs8_overview}
\end{figure*}

\subsection{Results on Linear Recurrence Models}
\label{sec:exp-end-to-end}

Table~\ref{tab:exp1_full_recompute_all} reports the end-to-end training throughput and memory usage for linear recurrence models across GBS 4, 8, and 16, while Figure~\ref{fig:exp1_gbs8_overview} visualizes the representative GBS\(=8\) results. For each workload, we report the best \(N\) that achieves the highest throughput. StateFlow improves throughput in every case, with speedups ranging from \(1.02\times\) to \(2.2\times\) and an average of
\(1.37\times\). It also reduces peak memory at every comparable point, with peak-memory reduction ranging from \(1.31\times\) to \(2.45\times\) with an average of \(1.63\times\), while enabling longer sequence configurations that run out of memory for the baseline method. Without activation recomputation, StateFlow achieves up to \(1.79\times\) throughput speedup and \(2.54\times\) peak-memory reduction on 3B GDN model.

\begin{table*}[!th]
    \centering
    \caption{
    Hybrid Period-4 activation recomputation results for
    (a) GDN and (b) Mamba-3.
    All entries report throughput/peak memory in K tokens/s and GB. Megatron and Swift denote the native pipelines, while ``+StateFlow'' uses the fastest feasible \((N,\alpha)\) configuration for each workload. OOM denotes out of memory.
    }
    \label{tab:exp2_hybrid_p4_full_recompute_all}

    \fontsize{9.1pt}{9.1pt}\selectfont
    \centerline{\textbf{(a) GDN}}
\begin{tblr}{
        width = \textwidth,
        colspec = {
            Q[c,m,wd=0.095\textwidth]
            *{9}{Q[c,m,wd=0.097\textwidth]}
        },
        rowsep = 0.8pt,
        colsep = 0.16pt,
        cells = {c,m},
        column{1} = {c,m},
        hline{1,2,3,7,8,12,13,Z} = {0.45pt, solid},
        vline{2,5,8} = {0.35pt, solid},
        row{1} = {
            font=\bfseries,
            abovesep=1.0pt,
            belowsep=1.0pt
        }
    }

    Method
    & GBS4 & GBS8 & GBS16
    & GBS4 & GBS8 & GBS16
    & GBS4 & GBS8 & GBS16 \\

    \SetRow{
        bg=modelgray,
        font=\bfseries,
        abovesep=0.8pt,
        belowsep=0.8pt
    }
    \SetCell{c,m} 3B
    & \SetCell[c=3]{c,m} 64K
    & &
    & \SetCell[c=3]{c,m} 128K
    & &
    & \SetCell[c=3]{c,m} 256K
    & & \\

    \SetRow{abovesep=1.2pt, belowsep=1.2pt}
    \SetCell{c,m} Megatron
    & 14.9/43.1G
    & 21.7/42.8G
    & 26.8/43.1G
    & 11.5/76.5G
    & 15.4/76.5G
    & 20.5/76.5G
    & OOM
    & OOM
    & OOM \\

    \SetRow{
        bg=sfblue,
        abovesep=1.2pt,
        belowsep=1.2pt
    }
    \SetCell{c,m,bg=sfblue}
    +StateFlow
    & \textbf{30.1}/29.3G
    & \textbf{30.5}/27.8G
    & \textbf{33.7}/35.9G
    & \textbf{25.5}/54.3G
    & \textbf{23.6}/42.5G
    & \textbf{26.7}/57.3G
    & \textbf{14.5}/76.4G
    & \textbf{14.8}/75.3G
    & \textbf{15.0}/76.1G \\

    \SetRow{abovesep=1.2pt, belowsep=1.2pt}
    \SetCell{c,m} Swift
    & 14.5/25.1G
    & 20.9/26.4G
    & 26.1/26.5G
    & 11.1/45.7G
    & 16.1/48.6G
    & 19.7/48.7G
    & OOM
    & OOM
    & OOM \\

    \SetRow{
        bg=sfblue,
        abovesep=1.2pt,
        belowsep=1.2pt
    }
    \SetCell{c,m,bg=sfblue}
    +StateFlow
    & \textbf{29.2}/20.1G
    & \textbf{30.9}/21.5G
    & \textbf{32.2}/21.5G
    & \textbf{24.1}/39.5G
    & \textbf{25.4}/39.5G
    & \textbf{26.0}/39.9G
    & \textbf{14.1}/64.4G
    & \textbf{14.4}/64.4G
    & \textbf{13.1}/65.6G \\

    \SetRow{
        bg=modelgray,
        font=\bfseries,
        abovesep=0.8pt,
        belowsep=0.8pt
    }
    \SetCell{c,m} 15B
    & \SetCell[c=3]{c,m} 64K
    & &
    & \SetCell[c=3]{c,m} 128K
    & &
    & \SetCell[c=3]{c,m} 256K
    & & \\

    \SetRow{abovesep=1.2pt, belowsep=1.2pt}
    \SetCell{c,m} Megatron
    & 10.5/49.2G
    & 13.1/49.2G
    & 15.2/49.2G
    & OOM
    & OOM
    & OOM
    & OOM
    & OOM
    & OOM \\

    \SetRow{
        bg=sfblue,
        abovesep=1.2pt,
        belowsep=1.2pt
    }
    \SetCell{c,m,bg=sfblue}
    +StateFlow
    & \textbf{14.1}/31.3G
    & \textbf{15.3}/30.8G
    & \textbf{15.9}/32.2G
    & \textbf{12.8}/45.2G
    & \textbf{13.2}/49.3G
    & \textbf{12.8}/43.6G
    & \textbf{9.8}/74.2G
    & \textbf{9.4}/77.1G
    & \textbf{10.1}/72.9G \\

    \SetRow{abovesep=1.2pt, belowsep=1.2pt}
    \SetCell{c,m} Swift
    & 11.1/44.8G
    & 14.0/46.7G
    & 16.0/49.4G
    & OOM
    & OOM
    & OOM
    & OOM
    & OOM
    & OOM \\

    \SetRow{
        bg=sfblue,
        abovesep=1.2pt,
        belowsep=1.2pt
    }
    \SetCell{c,m,bg=sfblue}
    +StateFlow
    & \textbf{15.3}/24.3G
    & \textbf{16.4}/25.0G
    & \textbf{17.0}/25.6G
    & \textbf{13.5}/37.0G
    & \textbf{13.9}/43.7G
    & \textbf{14.1}/36.6G
    & \textbf{9.2}/57.5G
    & \textbf{8.6}/65.8G
    & \textbf{10.2}/55.9G \\

    \SetRow{
        bg=modelgray,
        font=\bfseries,
        abovesep=0.8pt,
        belowsep=0.8pt
    }
    \SetCell{c,m} 32B
    & \SetCell[c=3]{c,m} 32K
    & &
    & \SetCell[c=3]{c,m} 64K
    & &
    & \SetCell[c=3]{c,m} 128K
    & & \\

    \SetRow{abovesep=1.2pt, belowsep=1.2pt}
    \SetCell{c,m} Megatron
    & 8.2/45.0G
    & 10.4/45.0G
    & 11.8/45.0G
    & 7.9/73.7G
    & 10.0/73.7G
    & 11.5/73.7G
    & OOM
    & OOM
    & OOM \\

    \SetRow{
        bg=sfblue,
        abovesep=1.2pt,
        belowsep=1.2pt
    }
    \SetCell{c,m,bg=sfblue}
    +StateFlow
    & \textbf{9.6}/34.5G
    & \textbf{11.4}/34.6G
    & \textbf{12.4}/34.7G
    & \textbf{10.7}/42.2G
    & \textbf{11.7}/42.8G
    & \textbf{12.1}/43.5G
    & \textbf{10.6}/56.6G
    & \textbf{10.2}/53.8G
    & \textbf{9.9}/56.1G \\

    \SetRow{abovesep=1.2pt, belowsep=1.2pt}
    \SetCell{c,m} Swift
    & 9.5/38.0G
    & 12.1/40.3G
    & 14.0/40.3G
    & 8.9/66.6G
    & 11.3/68.6G
    & 12.8/74.5G
    & OOM
    & OOM
    & OOM \\

    \SetRow{
        bg=sfblue,
        abovesep=1.2pt,
        belowsep=1.2pt
    }
    \SetCell{c,m,bg=sfblue}
    +StateFlow
    & \textbf{12.0}/22.9G
    & \textbf{13.0}/23.1G
    & \textbf{14.1}/27.4G
    & \textbf{11.8}/29.4G
    & \textbf{12.2}/28.1G
    & \textbf{12.8}/33.8G
    & \textbf{10.6}/46.7G
    & \textbf{11.0}/46.9G
    & \textbf{11.4}/42.6G \\

    \end{tblr}
    \centerline{\textbf{(b) Mamba-3}}
\begin{tblr}{
        width = \textwidth,
        colspec = {
            Q[c,m,wd=0.095\textwidth]
            *{9}{Q[c,m,wd=0.097\textwidth]}
        },
        rowsep = 0.8pt,
        colsep = 0.16pt,
        cells = {c,m},
        column{1} = {c,m},
        hline{1,2,3,7,8,12,13,Z} = {0.45pt, solid},
        vline{2,5,8} = {0.35pt, solid},
        row{1} = {
            font=\bfseries,
            abovesep=1.0pt,
            belowsep=1.0pt
        }
    }

    Method
    & GBS4 & GBS8 & GBS16
    & GBS4 & GBS8 & GBS16
    & GBS4 & GBS8 & GBS16 \\

    \SetRow{
        bg=modelgray,
        font=\bfseries,
        abovesep=0.8pt,
        belowsep=0.8pt
    }
    \SetCell{c,m} 3B
    & \SetCell[c=3]{c,m} 64K
    & &
    & \SetCell[c=3]{c,m} 128K
    & &
    & \SetCell[c=3]{c,m} 256K
    & & \\

    \SetRow{abovesep=1.2pt, belowsep=1.2pt}
    \SetCell{c,m} Megatron
    & 13.1/42.3G
    & 18.6/42.3G
    & 23.5/42.3G
    & OOM
    & OOM
    & OOM
    & OOM
    & OOM
    & OOM \\

    \SetRow{
        bg=sfblue,
        abovesep=1.2pt,
        belowsep=1.2pt
    }
    \SetCell{c,m,bg=sfblue}
    +StateFlow
    & \textbf{27.3}/25.3G
    & \textbf{28.4}/33.0G
    & \textbf{29.7}/33.6G
    & \textbf{22.9}/40.5G
    & \textbf{22.7}/56.4G
    & \textbf{23.2}/58.6G
    & \textbf{15.8}/72.7G
    & \textbf{14.0}/73.0G
    & \textbf{14.2}/73.0G \\

    \SetRow{abovesep=1.2pt, belowsep=1.2pt}
    \SetCell{c,m} Swift
    & 12.9/25.6G
    & 18.4/30.6G
    & 23.3/30.6G
    & 10.1/46.8G
    & 14.5/56.8G
    & 18.6/56.8G
    & OOM
    & OOM
    & OOM \\

    \SetRow{
        bg=sfblue,
        abovesep=1.2pt,
        belowsep=1.2pt
    }
    \SetCell{c,m,bg=sfblue}
    +StateFlow
    & \textbf{26.6}/19.7G
    & \textbf{27.3}/20.5G
    & \textbf{27.9}/21.3G
    & \textbf{22.2}/35.1G
    & \textbf{22.6}/35.2G
    & \textbf{22.4}/36.4G
    & \textbf{15.2}/65.4G
    & \textbf{13.6}/64.7G
    & \textbf{13.8}/64.8G \\

    \SetRow{
        bg=modelgray,
        font=\bfseries,
        abovesep=0.8pt,
        belowsep=0.8pt
    }
    \SetCell{c,m} 15B
    & \SetCell[c=3]{c,m} 64K
    & &
    & \SetCell[c=3]{c,m} 128K
    & &
    & \SetCell[c=3]{c,m} 256K
    & & \\

    \SetRow{abovesep=1.2pt, belowsep=1.2pt}
    \SetCell{c,m} Megatron
    & 8.3/52.1G
    & 10.6/52.2G
    & 12.2/52.2G
    & OOM
    & OOM
    & OOM
    & OOM
    & OOM
    & OOM \\

    \SetRow{
        bg=sfblue,
        abovesep=1.2pt,
        belowsep=1.2pt
    }
    \SetCell{c,m,bg=sfblue}
    +StateFlow
    & \textbf{11.0}/32.8G
    & \textbf{11.9}/32.5G
    & \textbf{12.5}/33.9G
    & \textbf{10.4}/46.3G
    & \textbf{10.6}/49.0G
    & \textbf{10.0}/45.4G
    & \textbf{8.4}/73.4G
    & \textbf{7.5}/73.5G
    & \textbf{8.6}/73.4G \\

    \SetRow{abovesep=1.2pt, belowsep=1.2pt}
    \SetCell{c,m} Swift
    & 8.9/48.7G
    & 11.2/51.3G
    & 12.9/54.1G
    & OOM
    & OOM
    & OOM
    & OOM
    & OOM
    & OOM \\

    \SetRow{
        bg=sfblue,
        abovesep=1.2pt,
        belowsep=1.2pt
    }
    \SetCell{c,m,bg=sfblue}
    +StateFlow
    & \textbf{12.6}/26.1G
    & \textbf{13.7}/27.2G
    & \textbf{14.3}/27.9G
    & \textbf{11.7}/38.2G
    & \textbf{11.9}/46.3G
    & \textbf{12.0}/37.2G
    & \textbf{8.2}/58.0G
    & \textbf{8.4}/58.2G
    & \textbf{9.1}/57.0G \\

    \SetRow{
        bg=modelgray,
        font=\bfseries,
        abovesep=0.8pt,
        belowsep=0.8pt
    }
    \SetCell{c,m} 32B
    & \SetCell[c=3]{c,m} 32K
    & &
    & \SetCell[c=3]{c,m} 64K
    & &
    & \SetCell[c=3]{c,m} 128K
    & & \\

    \SetRow{abovesep=1.2pt, belowsep=1.2pt}
    \SetCell{c,m} Megatron
    & 6.1/47.4G
    & 7.6/47.4G
    & 8.9/47.4G
    & OOM
    & OOM
    & OOM
    & OOM
    & OOM
    & OOM \\

    \SetRow{
        bg=sfblue,
        abovesep=1.2pt,
        belowsep=1.2pt
    }
    \SetCell{c,m,bg=sfblue}
    +StateFlow
    & \textbf{6.7}/30.2G
    & \textbf{8.1}/36.2G
    & \textbf{9.2}/47.1G
    & \textbf{7.8}/43.1G
    & \textbf{8.5}/43.1G
    & \textbf{8.5}/42.5G
    & \textbf{7.5}/55.2G
    & \textbf{7.7}/55.4G
    & \textbf{7.2}/52.0G \\

    \SetRow{abovesep=1.2pt, belowsep=1.2pt}
    \SetCell{c,m} Swift
    & 6.9/40.3G
    & 8.7/42.6G
    & 10.1/42.9G
    & OOM
    & OOM
    & OOM
    & OOM
    & OOM
    & OOM \\

    \SetRow{
        bg=sfblue,
        abovesep=1.2pt,
        belowsep=1.2pt
    }
    \SetCell{c,m,bg=sfblue}
    +StateFlow
    & \textbf{9.2}/22.9G
    & \textbf{9.8}/29.8G
    & \textbf{10.6}/24.9G
    & \textbf{9.1}/37.5G
    & \textbf{10.1}/38.7G
    & \textbf{10.1}/35.3G
    & \textbf{8.6}/48.3G
    & \textbf{8.9}/48.4G
    & \textbf{9.3}/43.6G\\

    \end{tblr}
\end{table*}

Figure~\ref{fig:exp1_n_sensitivity} shows that StateFlow selects PP8/TP1 in every case and is \(1.07\times\)--\(1.84\times\) faster than the best-tuned native topology. Throughput varies nonmonotonically with \(N\), as moderate chunking reduces pipeline bubbles, whereas overly fine chunking provides diminishing bubble
reductions and lowers kernel efficiency. Peak memory decreases and then plateaus as \(N\) increases.

StateFlow improves throughput across all GBS values. The gains are larger at
smaller GBS, where the native pipeline has fewer microbatches to amortize
pipeline bubbles. At larger GBS, StateFlow continues to improve throughput
while reducing peak memory. The smaller gains on the 15B and 32B models likely reflect heavier tensor-parallel communication and fewer baseline pipeline bubbles, leaving less overhead for StateFlow to remove.

StateFlow supports all of our evaluation configurations, whereas the native pipelines run out of memory in many long-context settings. The additional capacity results from releasing activations after chunk-level backward computation instead of retaining entire sequences. The improvement is consistent across GDN and Mamba-3. Although Swift is generally more memory-efficient than Megatron, StateFlow further reduces its peak memory, indicating that StateFlow's benefits are not tied to a specific training backend.

\begin{figure*}[!th]
    \centering
    \includegraphics[width=\textwidth]
    {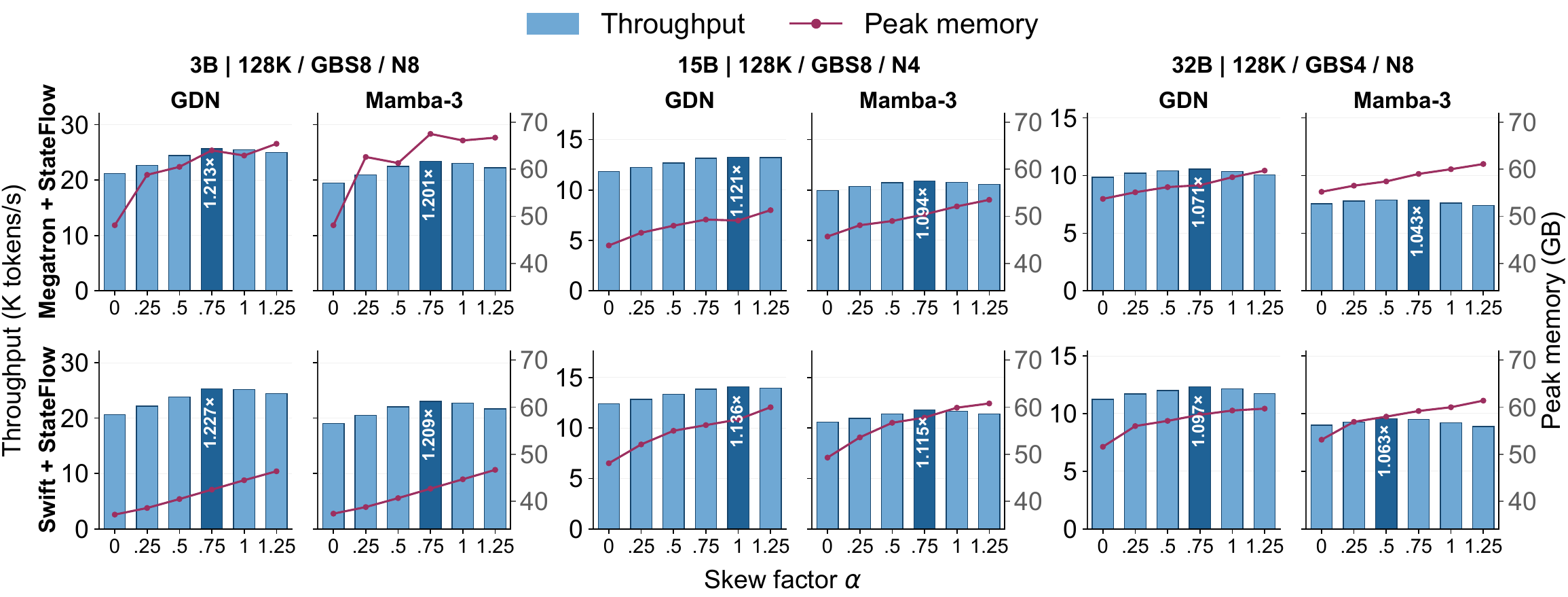}
    \caption{
    Sensitivity to the hybrid partition parameter \(\alpha\) on representative 128K context length. GBS and the chunk number \(N\) are fixed as labeled. Bars and lines denote throughput and peak memory, respectively. The dark bar marks the highest-throughput feasible \(\alpha\), annotated with its speedup over equal-length partitioning (\(\alpha=0\)). OOM denotes out of memory.
    }
    \Description{
    Twelve panels show throughput bars and peak-memory lines as \(\alpha\)
    varies for 3B, 15B, and 32B GDN and Mamba-3 models on Megatron and
    Swift. Dark bars indicate the highest-throughput feasible \(\alpha\).
    }
    \label{fig:exp2_hybrid_alpha}
\end{figure*}

\subsection{Results on Hybrid Models}
\label{sec:exp_hybrid}

Table~\ref{tab:exp2_hybrid_p4_full_recompute_all} reports the end-to-end
results for hybrid models. Each StateFlow entry uses the fastest feasible
\((N,\alpha)\) pair selected for that workload. Across
settings where the native pipeline completes, StateFlow matches or improves
throughput, achieving an average speedup of \(1.35\times\) and a maximum of
\(2.22\times\). It reduces peak memory by \(1.56\times\) on average and up to
\(2.44\times\). These benefits are consistent across GDN and Mamba-3 and
across Megatron and Swift frameworks.

The improvements persist as both model size and context length increase, and
each result includes inter-chunk state propagation and kernel-launch overhead.
Despite retaining additional K/V boundary buffers for softmax-attention layers,
StateFlow substantially expands the feasible workload range: it completes every
GBS setting at 256K for 3B and 15B and at 128K for 32B, where all native
pipelines run out of memory. 

Figure~\ref{fig:exp2_hybrid_alpha} isolates the effect of \(\alpha\) with
\(N\) fixed. For all four 3B configurations, \(\alpha=0.75\) achieves the
highest throughput, providing \(1.20\times\)--\(1.23\times\) speedups over
equal-length partitioning. For 15B and 32B, the selected \(\alpha\) ranges
from \(0.5\) to \(1.0\), with speedups of \(1.04\times\)--\(1.14\times\).
Throughput generally first increases and then decreases as the partition moves
from equal-length toward and beyond FLOP-balanced chunking. FLOP balance does not necessarily imply runtime balance because kernel
efficiency, memory behavior, and other system effects also vary with chunk
shape. Consequently, neither equal-length nor FLOP-balanced partitioning is
consistently optimal, demonstrating the value of profile-guided selection.
Although peak memory generally increases with \(\alpha\), all evaluated
configurations remain within the GPU memory limit.

\newcommand{\gdncfg}[1]{%
    {\fontsize{8pt}{8.4pt}\selectfont\texttt{#1}}%
}

\begin{table}[!th]
    \centering
    \caption{
        State transition overlap results. (a)~GDN speedup over the official implementation, with the ``\(B_V\)/split fraction''. (b)~Mamba-3 speedup on the selected overlap kernel region. Mamba-3's split fraction is always 0.5 and omitted.
    }
    \label{tab:state_path_optimization}

    \fontsize{9pt}{9pt}\selectfont
    \setlength{\tabcolsep}{1.8pt}

    \centerline{\textbf{(a) GDN speedup}}
    % \vspace{1pt}

    {
    \renewcommand{\arraystretch}{0.92}

    \begin{tabular*}{\columnwidth}{
        @{\extracolsep{\fill}}cccccc@{}
    }
        \toprule
        Model & Pass & 8K & 16K & 32K & 64K \\
        \midrule

        \multirow[c]{5}{*}{\shortstack[c]{3B\\PP8/TP1}}
        & \multirow[c]{2}{*}{Forward}
        & \(1.34\times\)
        & \(1.37\times\)
        & \(1.37\times\)
        & \(1.30\times\) \\[-1.6pt]

        &
        & \gdncfg{BV32/f.7}
        & \gdncfg{BV32/f.7}
        & \gdncfg{BV32/f.7}
        & \gdncfg{BV32/f.7} \\
        \cmidrule(lr){2-6}

        & \multirow[c]{2}{*}{Backward}
        & \(1.27\times\)
        & \(1.30\times\)
        & \(1.34\times\)
        & \(1.28\times\) \\[-1.6pt]

        &
        & \gdncfg{BV32/f.7}
        & \gdncfg{BV32/f.5}
        & \gdncfg{BV32/f.7}
        & \gdncfg{BV32/f.5} \\
        \cmidrule(lr){2-6}

        & Total
        & \(1.29\times\)
        & \(1.31\times\)
        & \(1.34\times\)
        & \(1.28\times\) \\
        \midrule

        \multirow[c]{5}{*}{\shortstack[c]{15B\\PP4/TP4}}
        & \multirow[c]{2}{*}{Forward}
        & \(1.37\times\)
        & \(1.37\times\)
        & \(1.39\times\)
        & \(1.35\times\) \\[-1.6pt]

        &
        & \gdncfg{BV16/f.3}
        & \gdncfg{BV16/f.9}
        & \gdncfg{BV16/f.9}
        & \gdncfg{BV16/f.5} \\
        \cmidrule(lr){2-6}

        & \multirow[c]{2}{*}{Backward}
        & \(1.35\times\)
        & \(1.37\times\)
        & \(1.42\times\)
        & \(1.42\times\) \\[-1.6pt]

        &
        & \gdncfg{BV32/f.5}
        & \gdncfg{BV32/f.7}
        & \gdncfg{BV32/f.5}
        & \gdncfg{BV32/f.6} \\
        \cmidrule(lr){2-6}

        & Total
        & \(1.35\times\)
        & \(1.37\times\)
        & \(1.41\times\)
        & \(1.40\times\) \\
        \midrule

        \multirow[c]{5}{*}{\shortstack[c]{32B\\PP4/TP8}}
        & \multirow[c]{2}{*}{Forward}
        & \(1.48\times\)
        & \(1.49\times\)
        & \(1.49\times\)
        & \(1.48\times\) \\[-1.6pt]

        &
        & \gdncfg{BV8/f.4}
        & \gdncfg{BV8/f.3}
        & \gdncfg{BV8/f.9}
        & \gdncfg{BV8/f.9} \\
        \cmidrule(lr){2-6}

        & \multirow[c]{2}{*}{Backward}
        & \(1.49\times\)
        & \(1.54\times\)
        & \(1.62\times\)
        & \(1.59\times\) \\[-1.6pt]

        &
        & \gdncfg{BV32/f.6}
        & \gdncfg{BV32/f.5}
        & \gdncfg{BV16/f.6}
        & \gdncfg{BV32/f.5} \\
        \cmidrule(lr){2-6}

        & Total
        & \(1.49\times\)
        & \(1.52\times\)
        & \({1.58\times}\)
        & \(1.56\times\) \\
        \bottomrule
    \end{tabular*}
    }

\centerline{\textbf{(b) Mamba-3 speedup}}
% \vspace{1pt}

{
\renewcommand{\arraystretch}{1.03}

\begin{tabular*}{\columnwidth}{
    @{\extracolsep{\fill}}cccccc@{}
}
    \toprule
    Model & Pass & 8K & 16K & 32K & 64K \\
    \midrule

    \multirow[c]{3}{*}{\shortstack[c]{3B\\PP8/TP1}}
    & Forward  & \(1.08\times\) & \(1.10\times\) & \(1.12\times\) & \(1.13\times\) \\
    & Backward & \(1.19\times\) & \(1.17\times\) & \(1.24\times\) & \(1.17\times\) \\
    & Total    & \(1.14\times\) & \(1.13\times\) & \(1.18\times\) & \(1.15\times\) \\
    \midrule

    \multirow[c]{3}{*}{\shortstack[c]{15B\\PP4/TP4}}
    & Forward  & \(1.30\times\) & \(1.30\times\) & \(1.27\times\) & \(1.21\times\) \\
    & Backward & \(1.52\times\) & \(1.54\times\) & \(1.64\times\) & \(1.53\times\) \\
    & Total    & \(1.39\times\) & \(1.40\times\) & \(1.41\times\) & \(1.34\times\) \\
    \midrule

    \multirow[c]{3}{*}{\shortstack[c]{32B\\PP4/TP8}}
    & Forward  & \(1.27\times\) & \(1.40\times\) & \(1.35\times\) & \(1.29\times\) \\
    & Backward & \(1.73\times\) & \(1.72\times\) & \(1.72\times\) & \(1.57\times\) \\
    & Total    & \(1.44\times\) & \(1.51\times\) & \(1.50\times\) & \(1.41\times\) \\
    \bottomrule
\end{tabular*}
}
\end{table}

\subsection{State Transition Grid Optimization and Overlap}
\label{sec:exp_state_passing_overlap}

Table~\ref{tab:state_path_optimization} summarizes the state transition optimization results for GDN and Mamba-3. Table~\ref{tab:state_path_optimization}~(a) reports GDN kernel speedups and the selected \(B_V\)/split fraction. Each forward and backward entry shows the speedup on the first line and the sweep-selected best configuration on the second. For example, \texttt{BV32/f.7} indicates \(B_V=32\) and a sequence split whose first segment contains \(70\%\) of the chunk.

The total GDN core speedup reaches \(1.34\times\), \(1.41\times\), and \(1.58\times\) for the 3B, 15B, and 32B configurations, respectively. The gains remain relatively stable from 8K to 64K, as state transition and the surrounding GDN computation scale similarly with sequence length. In the evaluated configurations, higher TP degrees reduce the per-rank state transition launch grid and its SM coverage, leaving more capacity for concurrent surrounding kernels. Backward generally favors larger \(B_V\) values, reflecting a different tradeoff between grid size and per-block efficiency. The selected split fractions also vary across passes and model configurations, motivating independent profiling rather than a fixed split.

Table~\ref{tab:state_path_optimization}~(b) reports Mamba-3 speedup on the overlap kernel regions defined by Equations~\ref{eq:fwd-split-schedule} and~\ref{eq:mamba3-bwd-overlap}. Forward and backward are tuned independently, and the \textit{Total} entry is computed from the summed forward and backward latencies before and after optimization.
% StateFlow accelerates Mamba-3's overlap-kernel regions by up to \(1.40\times\), \(1.74\times\), and \(1.46\times\) for forward, backward, and their total, respectively. 
StateFlow accelerates Mamba-3's overlapped regions by up to \(1.51\times\). This means that StateFlow hides up to \(35.7\%\), \(71.3\%\), and \(84.8\%\) of the fused Mamba-3 kernel latency from the critical path for the 3B, 15B, and 32B configurations, respectively.

The larger gains on the 15B and 32B configurations are consistent with their smaller fused-kernel launch grids: the 3B, 15B, and 32B configurations launch 80, 40, and 24 CTAs, respectively. Smaller grids leave more idle SM capacity for concurrent projection and weight gradient kernels. For forward, StateFlow profiles split points and the composition of the \(Out\) group. For 3B, \(Out\) contains the split output projection; for 15B and 32B, it additionally includes the MLP gate projection to better match the fused-kernel duration.

\section{Conclusion}

We have proposed StateFlow, an SPP system for training long-context linear recurrence and hybrid models. StateFlow uses chunks as pipeline units instead of full sequences, and uses compact boundary states to preserve forward/backward dependencies, enabling earlier activation release and finer-grained execution. It combines profile-guided hybrid partitioning with resource-aware state transition overlap to increase hardware utilization. 
% Together, these techniques make SPP practical across different recurrent operators, model scales, training frameworks, and recomputation settings.
Together, these techniques make StateFlow an efficient and scalable parallelism solution for training long-context models with linear recurrence.

\begingroup
\sloppy
\bibliographystyle{ACM-Reference-Format}
\bibliography{references}
\endgroup

\end{document}